\documentclass{article}

\usepackage{citesort}
\usepackage{epsfig}
\usepackage{amssymb}
\usepackage{amsmath}

\def\ud{\mathrm{d}}
\def\si{\sigma}
\def\de{\delta}
\def\De{\Delta}

\def\la{\langle}
\def\ra{\rangle}
\def\pa{\partial}
\def\fr{\frac}
\def\th{\theta}
\def\al{\alpha}

\begin{document}
\vspace*{1.0cm}
\noindent
{\bf
{\large
\begin{center}
On Peres' statement ``opposite momenta lead to opposite directions'', decaying systems and optical imaging
\end{center}
}
}

\vspace*{.5cm}
\begin{center}
W.\ Struyve, W.\ De Baere, J.\ De Neve and S.\ De Weirdt 
\end{center}

\begin{center}
Laboratory for Theoretical Physics\\
Unit for Subatomic and Radiation Physics\\
Proeftuinstraat 86, B--9000 Ghent, Belgium\\
E--mail: ward.struyve@ugent.be, willy.debaere@ugent.be 
\end{center}

\begin{abstract}
We re-examine Peres' statement ``opposite momenta lead to opposite directions''. It will be shown that Peres' statement is only valid in the large distance or large time limit. In the short distance or short time limit an additional deviation from perfect alignment occurs due to the uncertainty of the location of the source. This error contribution plays a major role in Popper's orginal experimental proposal. Peres' statement applies rather to the phenomenon of optical imaging, which was regarded by him as a verification of his statement. This is because this experiment can in a certain sense be seen as occurring in the large distance limit. We will also reconsider both experiments from the viewpoint of Bohmian mechanics. In Bohmian mechanics particles with exactly opposite momenta will move in opposite directions. In addition it will prove particularly usefull to use Bohmian mechanics because the Bohmian trajectories coincide with the conceptual trajectories drawn by Pittman {\em et al.} In this way Bohmian mechanics provides a theoretical basis for these conceptual trajectories.
\noindent

\end{abstract}

\section{Introduction}
Already in 1934 Popper \cite{popper34,popper82,tarozzi85pop} proposed an experiment which aimed to test the general validity of quantum mechanics. Popper assumes a source $S$ from which pairs of particles are emitted in opposite directions. Two observers, say Alice and Bob, are located at opposite sides of the source, both equiped with an array of detectors. If Alice puts a screen with a slit in her way of the particles, she will observe a diffraction pattern behind the screen. According to Popper, quantum mechanics will also predict a diffraction pattern on the other side of the source, where Bob is located, when coincidence counts are considered. This is because every measurement by Alice is in fact a virtual position measurement of the correlated particle on Bob's side, leading to an increased momentum uncertainty for Bob's particle as well. This is the same diffraction pattern that would be observed when a physical slit was placed on Bob's side. Popper, who declared himself a metaphysical realist, found this idea of ``virtual scattering'' absurd and predicted no increased momentum uncertainty for Bob's measurement due to Alice's position measurement. He therefore saw his proposed experiment as a possible test against quantum mechanics and in favour of his realist vision in which particles have at each time well defined positions and momenta; particles for which the Heisenberg uncertainty, for example, is only a lower, statistical limit of scatter.

Unfortunately, to describe the setup of the experiment, Popper occasionally invoked classical language, which veiled some severe problems which could obstruct a practical realization of his experiment (for an extensive discussion see Peres \cite{peres99}). Popper writes: ``We have a source $S$ (positronium, say) from which pairs of particles that have interacted are emitted in opposite directions. We consider pairs of particles that move in opposite directions \ldots''. It is the validity of this statement, which appears to be a very delicate issue, that we shall deal with in the present paper.

Of course, when we consider a decaying system at rest in classical mechanics, the fragments will have opposite momenta  
\begin{equation}
{\bf p}_1 + {\bf p}_2 = {\bf 0}
\label{1}
\end{equation}  
and if we take the place of decay of the system as the centre of our coordinate system the positions of the two fragments will satisfy $m_1 {\bf r}_{1}+ m_2 {\bf r}_{2} = {\bf 0}$, so the fragments will be found in opposite, isotropically distributed directions. If the fragments have equal masses then they will be found at opposite places, relative to the centre of the coordinate system. 

But these properties don't hold in quantum mechanics. Suppose that a system has a two--particle wavefunction $\psi$ which has a sharp distribution at ${\bf p}_1 + {\bf p}_2 = {\bf 0}$, i.e.\ both $|\la \hat{ p}_{1j}+\hat{ p}_{2j} \ra|$ and $\De (\hat{ p}_{1j}+\hat{ p}_{2j}) $ are small for every component $j$ of the momentum vectors $\hat{{\bf p}}_{1}+\hat{{\bf p}}_{2}$. Then according to the uncertainty relations 
\begin{equation}
\De (\hat{ p}_{1j}+\hat{ p}_{2j}) \De (m_1 \hat{ r}_{1j}+ m_2 \hat{ r}_{2j}) \ge \frac{\hbar}{2} (m_1 + m_2), 
\label{2}
\end{equation} 
the distribution of $m_1 r_{1j}+ m_2 r_{2j}$ will be broad for every component $j$. In the case of equal masses, the inequalities in Eq.~(\ref{2}) imply that opposite momenta are incompatible with opposite positions. In particular, at the moment of decay, the inequalities imply that opposite momenta of the particles are incompatible with the latter being both located at the origin of our coordinate system. It was even shown by Collett and Loudon \cite{collett872} that this initial uncertainty on the location of the source implies that Popper's experiment is inconclusive. 

Although the original proposal of Popper's experiment can hence not be performed practically, due to the fact that opposite momenta are incompatible with opposite positions, the intention of Popper's proposal can be maintained if we have a two--particle system which displays some form of entanglement in the position coordinates.{\footnote{The position entanglement should not be exact, otherwise, as shown by Short \cite{short00}, both of the observers would observe an infinite momentum spread.}} One is then in principle able to test experimentally wether one of the particles will experience an increased momentum spread due to a position measurement (within a slit width) of the correlated particle, i.e.\ we would be able to test a possible ``virtual scattering''. Such an experiment has actually been carried out by Kim and Shih \cite{kim99}, who used the phenomenon of optical imaging, which was reported by Pittman {\em et al.}\ \cite{pittman95}. In the process of optical imaging, position entangled photons are created by placing a lens in the way of one of the momentum entangled photons created by parametric down conversion.  

Recently Peres gave an analysis to which extent opposite momenta lead to opposite directions \cite{peres201}. He argues that the inequalities in Eq.~(\ref{2}) do not exclude a priori the possibility that opposite momenta of particles lead to opposite directions (instead of positions) where the particles will be found when a measurement is performed. On the contrary, the operator equivalent of (\ref{1}) would even lead to an observable alignment of the detection points of the two particles. However in section 2 we show that Peres' analysis can only be applied in what we will call the ``large time'' or ``large distance'' regime. This is the regime where the two particles have travelled a ``large'' distance from the source. In this limit the Heisenberg uncertainty will be of minor importance for the angular correlation. This result agrees with the scattering into cones theorem, which applies in the limit of infinite time. In the ``short'' distance regime the uncertainty on the location of the source, due to the Heisenberg uncertainty, leads to an additional error contribution to the angular alignment. We will also make clear what is exactly meant by the ``large'' or ``short'' distance regime. 

An example which shows the importance of the additional error, even in the long time regime,  is the orginial experiment of Popper (section 3). The situation is different for the phenomenon of optical imaging, which was considered by Peres as a justification for his analysis. Because the experiment can in a certain sense be seen as occuring in the long distance limit, Peres' analysis can be applied to this experiment.

In section 4, we will provide a simplified description of a decaying system using the Bohmian picture \cite{bohm1,bohm2}. It will prove usefull to use Bohmian mechanics because Bohmian mechanics inherits some of the language of classical mechanics. This is because Bohmian mechanics describes the motion of particles, allowing one to consider possible trajectories of the particles. This is contrary to ``standard quantum mechanics" where we can only speak of probabilities where a particle will be detected when a position measurement is performed. Before the position measurement is performed, particles can't, according to standard quantum mechanics, be seen as localized pointlike objects. As such, there is no basis in the standard quantum formalism to use trajectories of particles between two measurements. The classical language present in Bohmian mechanics can then be compared with the classical language used by Popper in his experimental proposal, in order to examine the possible interplay between opposite momenta and opposite positions.

We will also give an alternative explanation of the phenomenon of optical imaging (or at least of its massive particle equivalent) with the use of Bohmian mechanics (section 5). Although the experiment could be equally well described by quantum optics, there is an additional advantage by using Bohmian mechanics. The paper of Pittman {\em et al.}\ contains drawings of conceptual trajectories of the photons. These trajectories are not the real paths followed by the photons, as explained above, but merely serve as a tool to visualize the experiment. In section 4, we show that the Bohmian trajectories (calculated in the massive particle equivalent of the experiment of Pittman {\em et al.}), coincide with these conceptual trajectories in Pittman's paper. As such, Bohmian mechanics provides a theoretical basis for these trajectories. 

We want to stress that by using Bohmian mechanics we don't intend to contest the validity of standard quantum mechanics. Besides, as is well known Bohmian mechanics and standard quantum mechanics are completely equivalent at the empirical level. Now, wether one believes or not in the existence of particles as localized objects between two measurements, Bohmian mechanics provides a possible way of dealing with particle trajectories, consistent with quantum mechanics at the level of detections. Even if one is dissatisfied with the picture of moving particles, one can interpret the Bohmian trajectories as the flowlines of the probability, because the speeds of the Bohmian particles are defined as proportional to the quantum mechanical probability current. 

Finally we want to remark that, unfortunately Kim and Shih failed in their original intention to perform Popper's gedankenexperiment. It was shown by Short \cite{short00}, that due to imperfect momentum entanglement of the parametric down converted photons (caused by a restricted diameter of the source), the image was not perfect, which blurred the predicted results.   

\section{Opposite momenta and opposite directions} 
To discuss the possible angular alignment of momentum entangled particles, Peres considers a nonrelativistic wavefunction describing  massive particles. According to Peres, the reason for angular correlation of momentum entangled photons (as in the experiment of Pittman {\em et al.}) is the same as in the considered massive case.

The momentum correlated particles can be assumed to result from a decaying system at rest. The decaying system can then be described by the wavefunction
\begin{equation}
\psi({\bf r}_1,{\bf r}_2,t) = \int {F({\bf p}_1 , {\bf p}_2) e^{i({\bf p}_1 \cdot {\bf r}_1 +  {\bf p}_2  \cdot {\bf r}_2 - Et)/\hbar}  \ud {\bf p}_1  \ud {\bf p}_2  }
\label{3}
\end{equation}
where the momentum distribution $F$ is peaked around ${\bf p}_1 + {\bf p}_2 = {\bf 0}$ and around the rest energy $E_0$ of the decaying system. According to Peres the opposite momenta of the particles lead to opposite directions where the particles will be found when a measurement is performed. His argument goes as follows. The main contribution in the integral in Eq.~({\ref{3}}) comes from values ${\bf p}_1$ and ${\bf p}_2$ for which ${\bf p}_1 + {\bf p}_2 \simeq {\bf 0}$. Because of the rapid oscillations of the phase in the integrand in Eq.~({\ref{3}}), the integral will be appreciably different from zero only if the phase is stationary with respect to the six integration variables ${\bf p}_1$ and ${\bf p}_2$ in the vicinity of ${\bf p}_1 + {\bf p}_2 = {\bf 0}$, i.e.\ 
\begin{equation}
\fr{\pa S}{\pa {\bf p}_i} + {\bf r}_i - \fr{\pa E}{\pa  {\bf p}_i} t = {\bf 0}, \qquad  i=1,2
\label{4}
\end{equation}
where $S$ is the phase of $F({\bf p}_1 , {\bf p}_2)$ measured in units $\hbar$, i.e.\
\begin{equation} 
F({\bf p}_1 , {\bf p}_2) = |F({\bf p}_1 , {\bf p}_2)| e^{iS({\bf p}_1 , {\bf p}_2)/\hbar}
\label{5}
\end{equation}
and the equations ({\ref 4}) have to be evaluated for ${\bf p}_1 + {\bf p}_2 = {\bf 0}$. The equations ({\ref 4}) then determine the conditions on $ {\bf r}_i$ in order to have a non--zero $\psi$ (and $|\psi|^2$).

Peres then introduces spherical coordinates to describe ${\bf p}_i$ and ${\bf r}_i$, and varies the phase with respect to the six spherical variables $p_i,\phi_i,\th_i$ (the spherical coordinates of ${\bf p}_i$). Peres further assumes that the phase of $F$ obeys
\begin{equation}
\pa S/ \pa {\bf p}_i = {\bf 0}
\label{8}
\end{equation}
in the vicinity of ${\bf p}_1 + {\bf p}_2 = {\bf 0}$. This would restrict the place of decay near the origin of the coordinate system, because of Eq.~(\ref{4}). By varying with respect to the momentum angles, Peres obtains that the phase is stationary if ${\bf p}_i$ and ${\bf r}_i$ have the same direction. Because $F$ is peaked at ${\bf p}_1 + {\bf p}_2 = {\bf 0}$, this results in 
\begin{eqnarray}
\th'_1 + \th'_2 &=& \pi  \nonumber\\
|\phi'_1 -  \phi'_2| &=& \pi
\label{7}
\end{eqnarray}
where $r_i,\phi'_i,\th'_i$ are the spherical coordinates of ${\bf r}_i$. These equations show that the two particles can only be detected at opposite directions relative to the centre of our coordinate system. This is because the integral in Eq.~({\ref 3}) (and hence $|\psi|^2$) would only be appreciably different from zero if ${\bf r}_i$ obey Eq.~({\ref 7}). 

Peres mentions two error contributions to the angular alignment. The first is a transversal deviation of the order $\sqrt{ht/m}$ due to the spreading of the wavefunction, which was recognized as the standard quantum limit \cite{caves85}. The second is an angular spread of the order $\De (\hat{ p}_{1j}+\hat{ p}_{2j})/p_i$. One of the aims of the present paper is now to show that there is another error contribition which arises from the uncertainty on the source and which is particularly important in the ``short'' distance or ``short'' time regime. 

First we note that there is, apart from the conditions on $\th'_i$ and $\phi'_i$, also a condition on the variables $r_i$, which is not mentioned by Peres. This condition is obtained by varying the phase of the integrand with respect to $p_i$, having in mind the previous result that ${\bf p}_i$ and ${\bf r}_i$ have the same direction. If we define  $v_i  = \ud E / \ud p_i$, then the additional condition reads
\begin{equation}
r_i = v_i t 
\label{12}
\end{equation}
where the $v_i$ have to be evaluated for ${\bf p}_1 + {\bf p}_2 = {\bf 0}$. Because $\psi$ obeys the non-relativistic Schr\"odinger equation, $E$ equals $\frac{p^2_1}{2m_1} + \frac{p^2_2}{2m_2}$, with $m_i$ the masses of the particles. In this way Eq.~({\ref {12}})  becomes 
\begin{equation}
r_1 = \frac{p_1}{m_1} t, \quad r_2 = \frac{p_2}{m_2} t.
\label{9}
\end{equation}
Using ${\bf p}_1 + {\bf p}_2 = {\bf 0}$ one obtains
\begin{equation}
r_1 m_1 = r_2 m_2.
\label{13}
\end{equation}
Note that we haven't yet used the fact that $F$ is peaked around a certain energy $E_0$, as is required in the case of a decaying system at rest. As Peres remarks in his paper, a restriction of the energy to $E_0$ further restricts the momenta of the particles to satisfy $p^2_1 = p^2_2 = 2 E_0 m_1 m_2/(m_1 + m_2)$.

Combining (\ref{7}) and (\ref{13}) one obtains that the joint detection probability has a maximum for the classically expected relation $m_1 {\bf r}_1 + m_2 {\bf r}_2 = {\bf 0}$. Hence Peres' statement ``opposite momenta lead to opposite directions'' may be replaced by a stronger statement, i.e. the opposite momenta lead to a maximum detection probability for $m_1 {\bf r}_1 + m_2 {\bf r}_2 = {\bf 0}$. However we still have to consider the possible sources of deviation from this classical relation. Classically one can, in theory, make both quantities $\De (\hat{ p}_{1j}+\hat{ p}_{2j})$ and $\De (m_1 \hat{ r}_{1j} + m_2 \hat{ r}_{2j})$ as small as wanted. Quantum mechanically one can at best prepare the system, such that initially the equality in
\begin{equation}
\De (\hat{ p}_{1j}+\hat{ p}_{2j}) \De (m_1 \hat{ r}_{1j} + m_2 \hat{ r}_{2j}) \ge \frac{\hbar}{2} (m_1 + m_2 )
\label{14.11}
\end{equation} 
is reached. Remark that this equation implies that in case of opposite momenta, the particles can't depart from a confined, fixed source. 

Because the operator $\hat{ p}_{1j} + \hat{ p}_{2j}$ commutes with the free Hamiltonian, the variance of the momentum operator $\hat{ p}_{1j}+\hat{ p}_{2j}$ is stationary. The variance of $\hat{ r}_{1j}+\hat{ r}_{2j}$ however, will in general increase with time due to the spreading of the wavefunction. This can be seen if we write down the expression for the free evolution of the operator $m_1 \hat{ {\bf r}}_{1} + m_2 \hat{ {\bf r}}_{2}$ in the Heisenberg picture
\begin{equation}
m_1 \hat{ {\bf r}}_{1}(t) + m_2 \hat{{\bf r}}_{2}(t) = m_1 \hat{{\bf r}}_{1}(0) + \hat{{\bf p}}_{1}(0)t + m_2 \hat{ {\bf r}}_{2}(0) +\hat{{\bf p}}_{2}(0)t
\label{20}
\end{equation}
The variance of this operator for an arbitrary component $j$ is
\begin{eqnarray}
\De \big( m_1 \hat{ r}_{1j}(t) + m_2 \hat{ r}_{2j}(t) \big)^2 &=& \De \big( m_1 \hat{ r}_{1j}(0) + m_2 \hat{ r}_{2j}(0) \big)^2 + \De \big( \hat{ p}_{1j}(0)+\hat{ p}_{2j}(0) \big)^2 t^2\nonumber\\
&&+ \Big< \Big\{ \big( m_1 \hat{ r}_{1j}(0) + m_2 \hat{ r}_{2j}(0) \big), \big( \hat{ p}_{1j}(0)+\hat{ p}_{2j}(0) \big) \Big\} \Big> t \nonumber\\
&&- 2\big< m_1 \hat{ r}_{1j}(0) + m_2 \hat{ r}_{2j}(0)  \big> \big<  \hat{ p}_{1j}(0)+\hat{ p}_{2j}(0) \big>t
\label{21}
\end{eqnarray}
where the brackets $\{,\}$ denote the anticommutator. If we assume a distribution $F$ which is real and symmetric, i.e.\ $F({\bf p}_1 , {\bf p}_2) = F(-{\bf p}_1 , -{\bf p}_2)$ then the last two terms in Eq.~({\ref {21}}) are each zero. So the variance of $ m_1 \hat{ r}_{1j}(t) + m_2 \hat{ r}_{2j}(t) $ increases with time
\begin{equation}
\De \big( m_1 \hat{ r}_{1j}(t) + m_2 \hat{ r}_{2j}(t) \big)^2  =  \De \big( m_1 \hat{ r}_{1j}(0) + m_2 \hat{ r}_{2j}(0) \big)^2 + \De \big( \hat{ p}_{1j}(0)+\hat{ p}_{2j}(0) \big)^2 t^2 .
\label{22.1}
\end{equation}
This leads to an increasing deviation from the relation $m_1 {\bf r}_1 + m_2 {\bf r}_2 = {\bf 0}$. 

There is thus always an interplay between opposite momenta and opposite directions which is expressed in Eq.~(\ref{14.11}) and Eq.~(\ref{22.1}). We argue that one should study Eq.~(\ref{22.1}), where the variances at $t=0$ in the right hand side of the expression are limited by the Heisenberg uncertainty in Eq.~(\ref{14.11}), to determine to which extent we can speak of possible angular alignment. 

Let us now see how Peres' error contributions come about and in which regime they are important. Assume for convenience that $m_1 = m_2=m$. The transversal deviation $L(t)$ can be taken of the order $\De \big( \hat{ r}_{1j}(t) +  \hat{ r}_{2j}(t) \big)$. There are two contributions to this transversal deviation. The first is $\De \big( \hat{ r}_{1j}(0) +  \hat{ r}_{2j}(0) \big) = L(0)$ and is important for small times. The second contribution is $\De \big( \hat{ p}_{1j}+\hat{ p}_{2j} \big) t/m$ which becomes important for larger times. The angular deviation $\theta$ may be derived from $\tan(\theta) = L(t)/R(t)$, where $R(t)= pt/m$ is the distance that both particles have travelled. For small times one has $\tan(\theta) \simeq \De \big( \hat{ r}_{1j}(0) +  \hat{ r}_{2j}(0) \big)m/pt$ and for large times one has $\tan(\theta) \simeq \De \big( \hat{ p}_{1j} +  \hat{ p}_{2j} \big)/p$. Hence for large times we obtain the error mentioned by Peres. From the relations (\ref{14.11}) and (\ref{22.1}) one can also easily derive the standard quantum limit
\begin{eqnarray}
\De \big( \hat{ r}_{1j}(t) +  \hat{ r}_{2j}(t) \big)^2 &\ge& 2 \De \big(  \hat{ r}_{1j}(0)+ \hat{ r}_{2j}(0) \big)   \De \big( \hat{ p}_{1j}(0)+\hat{ p}_{2j}(0) \big) t/m  \nonumber\\
 &\ge& 2\hbar t/m.
\label{22.2}
\end{eqnarray}
However, this uncertainty is misleading because for small times it neglects the contribution arising from the uncertainty on the source $\De \big( m_1 \hat{ r}_{1j}(0)+ m_2 \hat{ r}_{2j}(0) \big)$. Especially in the considered case this contribution will be large because $\De \big( \hat{ p}_{1j}(0)+\hat{ p}_{2j}(0) \big)$ is small. 

In summary we see that Peres gave error contributions which only apply in the large time regime. These error contributions are in perfect agreement with the ``scattering into cones'' theorem which states that for every cone $C$ in $\mathbb{R}^m$ with apex in the origin 
\begin{equation}
\lim_{t \to \infty} \int_C d^m x |\psi(x,t)|^2 = \int_C   d^m p |\phi(p)|^2
\label{22.201}
\end{equation}
with $\phi(p)$ the momentum wave function \cite{newton}. This means that the probability that in the infinite future the particles will be found in the cone $C$ is equal to the probability that their momenta lie in the same cone. In the small time regime the uncertainty on the source gives the major contribution. We could say that the transition between the small time regime and the large time regime occurs at time $T$ when both contributions to $L(t)$ are equally large, i.e.\ $L(0) = \De \big( \hat{ p}_{1j}(0)+\hat{ p}_{2j}(0) \big) T/m$. Because we can at best reach the Heisenberg uncertainty, a lower bound for the time is given by $T= L(0)^2 k_c / c$ with $k_c$ the wavevector corresponding to the compton wavelength of the particles. In the next section we will plug some numbers into this relation to show that we are not making a trivial point in stressing this additional error contribution. Let it for the moment be sufficient to provide an example which illustrates the importance of this uncertainty on the source. We can use a Gaussian distribution to represent the momentum correlation
\begin{equation}
F({\bf p}_1 , {\bf p}_2) \sim e^{-\frac{({\bf p}_1 + {\bf p}_2)^2}{\si}}.
\label{10}
\end{equation}
The smaller the value of $\si$, the better the momentum correlation between the two fragments. In the limit $\si \to 0$ this distribution approaches the Dirac delta distribution. Remark that this distribution is not peaked around a certain energy $E_0$ as should be required for a decaying system at rest. In the Appendix it is explained why we can leave this restriction on the energy aside without changing the main result. It will follow that a reasonable energy width, peaked around $E_0$, will imply only a minor broadening of the wavefunction. The wavefunction at $t=0$ is
\begin{equation}
\psi({\bf r}_1 , {\bf r}_2,0) \sim \de({\bf r}_1 - {\bf r}_2) e^{-{\bf r}^2_1 \si /4 \hbar^2}.
\label{11}
\end{equation}   
Thus clearly $\psi$ represents a decaying system because initially ${\bf r}_1 = {\bf r}_2$. But for small values of $\si$ (when $F$ is peaked around ${\bf p}_1 + {\bf p}_2 = {\bf 0}$) the probability of finding the particles at $t=0$ at some configuration is totally smeared out, although the probability has a maximum at the origin of the coordinate system. Note that although the relation $\pa S/ \pa {\bf p}_i = {\bf 0}$ is satisfied, this condition doesn't restrict the place of decay near the origin of the coordinate system, as was assumed by Peres. So in the short time regime, it may be hard to speak of possible opposite movements of particles relative to the origin because we cannot exactly say (at least without measurement) where the decay of the system took place. 
As time increases the deviation from the detection probability peak at $m_1 {\bf r}_1 + m_2 {\bf r}_2 = {\bf 0}$ will even increase with time as was shown above. However, because the distance from the particles to the source increases, the uncertainty of the source will become less important for the angular deviation. In the long distance regime the angular deviation will then be dominated by the momentum uncertainty.

In section 4 we will show that we can retain the classical picture of a decaying system in the previous example, in both the short time regime and the long time regime, when it is described by Bohmian mechanics. Namely in the case of the considered wavefunction, the Bohmian particles will depart near eachother and will move along opposite directions. However, the place of departure will vary from pair to pair over an extended region. The more this initial region is confined, the less perfect the momentum entanglement will be, and the less perfect the opposite movements of the Bohmian particles will be.

\section{Opposite momenta and opposite directions in the case of the experiments of Popper and Pittman et al.}
The importance of the error contribution arising from the uncertainty of the source becomes clear if we consider the example of Popper's experimental proposal. It can easily be seen that Popper's experiment cannot occur in the large time regime. If the particles would travel long distances (and hence obtain good angular correlation), the virtual slit (which is of the order of the transversal deviation) would be too large to have virtual diffraction of the particles on Bob's side. Hence Popper's original experimental proposal should be considered in the short time regime. However, in this regime the uncertainty on the source becomes the most important contribution to the angular deviation and as follows from the analysis of Collett and Loudon, this uncertainty makes a detectable virtual diffraction impossible. This also implies that in discussing Popper's experiment one should be carefull with statements such as ``\ldots the allowed deviation from perfect alignment is of the order of $\De |p_1+ p_2|/|p_1 - p_2|$, which is much too small to be of any consequence in the present discussion.'' and ``\ldots nearly perfect alignment can be taken for granted, \ldots" \cite{peres99}.

In the case of optical imaging, performed by Pittman {\em et al.}, Peres' analysis can be applied. Let us first briefly review this experiment. The experiment uses momentum correlated photons resulting from spontaneous parametric down conversion (SPDC). In this process a pump photon incident on a nonlinear beta barium borate (BBO) crystal leads to the creation of a signal and idler photon. The place of creation  within the BBO crystal is unknown. Due to momentum conservation the sum of the momenta of these photons has to equal the momentum of the pump photon.  This results in the momentum entanglement of the two photons, because the momenta of the idler and signal photon can be combined in an infinite number of ways to equal the momentum of the pump photon. In the same way the energies of the created photons add up to the energy of the pump photon. In the experiment the signal and idler photons are sent in two different directions where coincidence records may be performed by two photon counting detectors. A convex lens, with focal length $f$, is placed in the signal beam in order to turn the momentum correlation of the created photons into spatial correlation (see Fig.\ 1). 
\begin{figure}[t]
\begin{center}
\epsfig{file=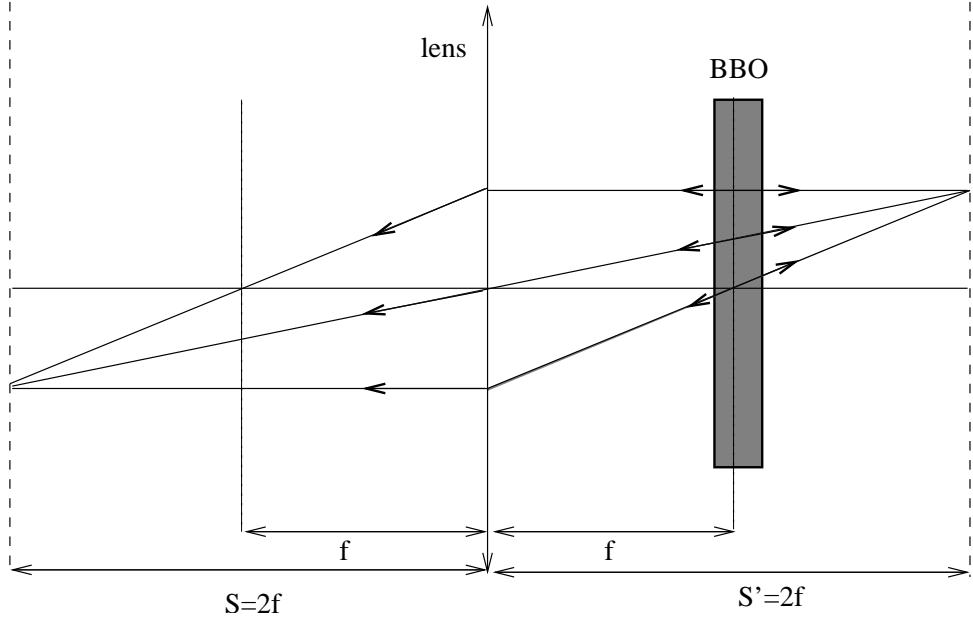}
\end{center}
\caption{Unfolded schematic of the experiment. Particle $1$ (the signal photon) is moving to the right and particle $2$ (the idler photon) is moving to the left. We have put $S=S'=2f$. Similar schematics are shown in \cite{pittman95}. There the trajectories represent conceptual photon trajectories, here the trajectories are the Bohmian trajectories of massive particles. The place of creation of the photons within the BBO crystal should be taken sufficiently extended in order to preserve the opposite momenta correlation.}
\label{bbo}
\end{figure}   
\noindent
In front of the detector for the signal beam an aperture is placed at a distance $S$ from the lens. By placing the detector for the idler beam at a distance $S'$ from the lens, prescribed by the Gaussian thin lens equation, i.e.\
\begin{equation}
\frac{1}{S} + \frac{1}{S'} = \frac{1}{f}
\label{23}
\end{equation}
and scanning in the transverse plane of the idler beam, an image of this aperture is observed in the coincidence counts. This image obeys the classical lens equations in the following sense. If a classical pointlike light source would be placed in the plane of the aperture, where the signal photon was detected, it would have an image where the idler photon was detected. This is the spatial correlation of the photons.

The experiment can be seen to occur in the long distance regime because of the lens. In some sense the lens can be seen as projecting the angular correlation at infinity, to finite distances (at distances $2f$ from the lens). The better the momentum correlation, the better the angular correlation is at infinity or the better the optical imaging is. 

It is interesting to use the data of this experiment to give an example where the border can be situated between the long distance and the short distance regime for a stongly momentum correlated system. A lower bound for the time is $T= L(0)^2 k_c / c$, hence if there would be no lens in the Pittman experiment, the distance travelled by the photons at that time would be $R=  L(0)^2 k_c \simeq 70m $ ($L(0)= 2mm$ is the width of the pump beam and the wavelength of the pump photon is $351.1nm$). This means that if we would create momentum entangled photons via SPDC, then the angular deviation would be dominated by the error arising from the uncertainty of the source within a distance of $R$.

\section{Bohmian picture of a decaying system}
In this section we give the Bohmian description of a system consisting of two particles with opposite momenta, resulting from a decaying system at rest.{\footnote{Our simplified Bohmian approach to a decaying system is to be distinguished from the one studied by Y.\ Nogami, F.M.\ Toyama and W. van Dijk \cite{nogami00}, where a decaying system is represented by a particle that leaks out from a region surrounded by a repulsive potential barrier.}} We use the wavefunction in Eq.~({\ref{3}}) with momentum distribution 
\begin{equation}
F({\bf p}_1 , {\bf p}_2) = N \de ({\bf p}_1 + {\bf p}_2)  e^{-\alpha p^2_1 / \hbar}
\label{24}
\end{equation}
where $N$ is a normalisation factor. The parameter $\al$ sets the scale of the initial separation of the two particles, as will be seen soon. A small $\al$ will correspond to the considered physical situation, i.e.\ a decaying system. The parameter is introduced in order to avoid singularities arising from the delta distribution when calculating the Bohmian trajectories later on. Again we have left aside the restriction on the total energy, as required in the case of a decaying system (see Appendix). Remark that the exponential factor in Eq.~(\ref{24}) doesn't restrict the value of the total energy for a small $\al$. The wavefunction corresponding to the distribution $F$ is
\begin{equation}
\psi({\bf r}_1 , {\bf r}_2,t) = N \bigg(\frac{\pi \hbar}{\al + it/2\mu} \bigg)^{3/2} e^{-({\bf r}_1 - {\bf r}_2)^2 / 4 \hbar (\al + \frac{it}{2\mu} )}
\label{25}
\end{equation}
with $\mu$ the reduced mass of the fragments: $\frac{1}{\mu} = \frac{1}{m_1} +\frac{1}{m_2}$.
At $t=0$ the probability distribution is
\begin{equation}
|\psi({\bf r}_1 , {\bf r}_2,0)|^2 = N^2 \bigg(\frac{ \pi \hbar}{\al}\bigg)^3  e^{-({\bf r}_1 - {\bf r}_2)^2 / 2 \hbar \al}.
\label{26}
\end{equation}
It follows that a small value for $\al$ corresponds to the considered physical situation of a decaying system. However, the place of decay is unknown. This is a consequence of the Heisenberg uncertainty, as explained in the section 2. 

The Bohmian trajectories ${\bf R}_j(t)$ of the particles are found by solving the differential equations  
\begin{equation}
\frac{\ud {\bf R}_j}{\ud t} = \frac{1}{m_j} \frac{\textrm{Re} \psi^*({\bf r}_1 , {\bf r}_2,t) {\hat {\bf p}}_j \psi({\bf r}_1 , {\bf r}_2,t)}{|\psi({\bf r}_1 , {\bf r_2},t)|^2 } \Bigg|_{{\bf r}_i = {\bf R}_i}.
\label{27}
\end{equation}
Because $(\hat {\bf p}_1 + \hat {\bf p}_2) \psi = 0$ the trajectories of the particles satisfy
\begin{equation}
\frac{\ud}{\ud t} (m_1 {\bf R}_1 + m_2 {\bf R}_2) = {\bf 0}.
\label{28}
\end{equation}
This shows that the particles have opposite speeds and thus move in opposite directions. Integration of the differential equations (\ref{27}) leads to
\begin{eqnarray}
{\bf R}_1(t) &=& {\bf C}_1 + {\bf C}_2 \sqrt{t^2/4\mu^2 + \al^2} \nonumber\\
{\bf R}_2(t) &=& {\bf C}_1 - {\bf C}_2 \sqrt{t^2/4\mu^2 + \al^2}
\label{29}
\end{eqnarray}
where ${\bf C}_1$ and ${\bf C}_2$ are arbitrary constant vectors. It follows that the particles also move along straight lines. If we consider an ensemble of identically prepared systems, the probability density of the Bohmian particles equals the quantum mechanical distribution $|\psi|^2$ \cite{bohm1,bohm2}. This is also called the quantum equilibrium hypothesis \cite{durr} and ensures the empirical equivalence between Bohmian mechanics and standard quantum mechanics. This equality determines the distribution of the vectors ${\bf C}_i$ in the ensemble. Because the probability distribution at $t=0$ is sharply peaked at ${\bf r}_1 = {\bf r}_2$ (for small values of $\al$), the particles depart near each other. As follows from Eq.~(\ref{29}) their further propagation proceeds along straight lines, in the direction of their connecting line. Thus opposite momenta lead to opposite directions of movement for Bohmian particles. But their place of departure is located within an extended area, in order to preserve momentum correlation. 

By using Bohmian mechanics, we are thus able to retain part of the classical picture of a decaying system at rest. Remark the similarity in language with the one used by Popper to describe his experiment. The difference is that Popper assumed the particles to depart from a confined region (which is however incompatible with opposite momenta in quantum mechanics). Note that the opposite movements of Bohmian particles can not be verified experimentally. This is because a measurement of the place of departure of the Bohmian particles would cause a change of the wavefunction (this is the collapse of the wavefunction in standard quantum mechanics), which in turn implies different particle trajectories.

\section{The experiment of Pittman {\em et al.}\ revisited by Bohmian mechanics}
Although the experiment of Pittman {\em et al.}\ can be correctly explained with quantum optics, we will provide a Bohmian account of the experiment, when it is ``translated'' into its massive particle equivalent. One of the reasons to use Bohmian mechanics is that it justifies the conceptual photon trajectories drawn by Pittman {\em et al.}\  \cite{pittman95}. I.e.\ the photon trajectories coincide with the trajectories of Bohmian particles in the massive particle equivalent of the experiment. This (Bohmian) quantum mechanical approach is contrary to the explanation in terms of ``usual'' geometrical optics used by Pittman {\em et al.} In quantum optics these paths are usually regarded as a visualisation of the different contributions to the detection probabilities.  

Because there is at present no satisfactory Bohmian particle interpretation for photons, see discussion elsewhere \cite{holland93,bohm5,kaloyerou94,struyve031}, we will follow Peres' departure and consider the nonrelativistic massive particle wavefunction in Eq.~(\ref{3}) instead of quantum optics, to study the experiment by Pittman {\em et al.} The SPDC source then corresponds to a decaying system at rest, resulting in two energy and momentum correlated fragments. We will assume the total momentum of the fragments to be zero, instead of some fixed value corresponding to the initial momentum of the total system (which would represent the momentum of the pump photon). This assumption corresponds to the ``unfolded'' schematic introduced by Pittman {\em et al.}\ \cite{pittman95} (see Fig.\ 1). In this way we can use the momentum distribution $F$ defined in previous section
\begin{equation}
F({\bf p}_1 , {\bf p}_2) = N \de ({\bf p}_1 + {\bf p}_2)  e^{-\alpha p^2_1 / \hbar}.
\label{24.0000000003}
\end{equation}
In previous section we described the free evolution after the decay of the system. The unknown place of decay in the massive particle case corresponds to the unknown place of creation within the BBO crystal in the photon case (which is also related to the width of the pump beam). To complete the Bohmian description of the massive particle equivalent of optical imaging, we just have to describe the system's interaction with the lens. In classical optics we can use ray optics to describe the action of the lens on an impinging light beam \cite{bornwolf80}. The rays are such that the Gaussian thin lens equations are satisfied. Two generic examples, which we will need later on, are the following. The effect of the lens on a plane wave is to turn it into a converging wave, with focus in the focal plane, such that the corresponding rays obey the lens equations. The characteristics of the converging wave are then determined by the momentum of the incoming plane wave and the focal length. A second example is a spherical wave, representing a point source. If we assume that the light source is located in a plane at a distance $S$ from the lens, then the spherical wave will turn into a converging wave with focus in the plane at a distance $S'$ from the lens so that $1/S + 1/S' = 1/f$ and the source, the image and the centre of the lens will be aligned. In massive particle quantum physics the equivalent of optical lenses are electrostatic or magnetic lenses. These electromagnetic lenses are generally used to collimate or focus beams of charged particles. This field of research is usually called optics of charged--particle beams or the theory of charged--particle beams through electromagnetic systems. Most of the literature deals with the classical description of the particles and only recently the quantum mechanical approach has been studied, see for example Hawkes and Kasper \cite{hawkes96}, and Khan and Jagannathan \cite{khan95} and references therein. Here we will not bother about the detailed analysis of particles passing through such electromagnetic lenses, and use directly, in the spirit of de Broglie, the analogy with classical optics. For example we can describe the action of an electromagnetic lens as turning a quantum mechanical plane wave into a Gaussian wave (we can take this as the analogue of the converging wave in classical optics, because a Gaussian wave is contracting before expanding), determined by the momentum of the incoming wave and the focal length. This analogy is very appealing because the rays in classical optics can be ``identified'' with the Bohmian trajectories. This is because in the one--particle case, the curves determined by the normals of the wavefronts of the quantum mechanical wavefunction are just the possible orbits of a particle described by Bohmian mechanics \cite{holland}. If we apply this to our decaying system, then every plane wave of the particle impinging on the lens, say particle two, in the integral in Eq.~(\ref{3}) is turned into a particular Gaussian wave. The resulting wave is then
\begin{equation}
\psi'({\bf r}_1,{\bf r}_2,t) = \int {F({\bf p}_1 , {\bf p}_2) e^{i({\bf p}_1 \cdot {\bf r}_1  - p^2_1t/2m_1)/\hbar}  } G({\bf r}_2,{\bf p}_2,f)\ud {\bf p}_1  \ud {\bf p}_2  
\label{30}
\end{equation}
where $G$ represents the Gaussian wave. This wave is guiding the particles after particle two passed the lens. To avoid unnecessary mathematical complications when calculating the Bohmian trajectories implied by the wave Eq.~(\ref{30}), we assume that the place of decay of the system is somewhere in the middle between the lens and the detector on the right (where the idler photon arrives in the experiment of Pittman {\em et.\ al.}). This corresponds to a BBO crystal placed in the middle instead of it placed near the lens, as in the experiment. When particle two arrives in the vicinity of the lens, particle one arrives in the vicinity of the detector placed on the right. We can describe the effect of the detector on the wavefunction as the first stage of a von Neumann measurement process (see for example Bohm and Hiley \cite{bohm84}). In this process the wavefunction gets entangled with a pointer which is represented by a wavefunction $\phi$ with a small width. In this way the wavefunction in Eq.~(\ref{3}) turns into
\begin{equation}
\psi''({\bf r}_1,{\bf r}_2,{\bf y},t)= \int {F({\bf p}_1 , {\bf p}_2) \delta({\bf a} - {\bf r}_1) \phi({\bf y},t,{\bf a}) }  e^{i\big( {\bf p}_1 \cdot ({\bf a} - {\bf r}_2)  - Et\big)/\hbar} \ud {\bf p}_1  \ud {\bf p}_2 \ud {\bf a}  
\label{31}
\end{equation}
where we have written the factorization into position eigenfunctions $\delta({\bf a} - {\bf r}_1)$ of particle one explicitly. It is supposed that the interaction of the detecting apparatus with the decaying system lasted long enough in order to assure that the wavepackets $\phi({\bf y},{\bf a})$ are non overlapping for each ${\bf a}$. During this process, the pointer particle has entered one of the packets $\phi({\bf y},{\bf a})$, determined by its initial position. While the packets are non overlapping, only the considered wavepacket is determining the subsequent trajectory of the pointer particle. The other wavepackets can then be dismissed for the further description of the system. In conventional quantum mechanics this process would be treated as a collapse of the wavefunction, which is the second stage of the von Neumann measurement process. As a result, the effective wave guiding particle two, is also reduced to the following superposition of plane waves
\begin{eqnarray}
\!\!\!\!\psi_2({\bf r}_2,t)\!\! &=&\!\! N \int { e^{\frac{i}{\hbar}\big({\bf p} \cdot ({\bf a} -  {\bf r}_2) -   {p}^2 t/2m_2  \big) -\al p^2/\hbar} \ud {\bf p}  }\nonumber\\ 
&=&\!\! N \bigg(\frac{\pi \hbar}{\al + it/2m_2} \bigg)^{3/2} e^{-({\bf a} - {\bf r}_2)^2 / 4 \hbar (\al + \frac{it}{2m_2} )}.
\label{32}
\end{eqnarray}
The phase of this wave is 
\begin{equation}
S({\bf r}_2,t) = \frac{t({\bf a} - {\bf r}_2)^2}{8m_2 \al^2 + t^2/m_2} - \frac{3 \hbar}{2} \tan^{-1} (t/2m_2 \alpha).
\label{33}
\end{equation}
So the wavefronts of the guiding wave of particle two are spheres with centre in ${\bf a}$. Because the detectors are placed in planes at distances $S$ and $S'$ from the lens, with $S$ and $S'$ obeying the Gaussian lens equation (\ref{23}), this wave will result, after propagation through the lens, in a converging wave with focus in the plane at a distance $S$ from the lens and where the focus is determined by the Gaussian thin lens equations. Hereby we used again the analogy with classical Gaussian optics. If for example $S=S'$ and if the centre of the lens is taken as the origin of our coordinate system, then the focus will be at $-{\bf a}$ (see Fig.\ 1). As a consequence of the quantum equilibrium hypothesis, particle two will be detected in the focus of the wave. Because we used a Gaussian to describe the converging wave, the Bohmian trajectories will not be straight lines, but will be curved (for images see Holland \cite{holland}). The curvature will depend on the width in the focus of the Gaussian. In the limit of a zero width, however, the trajectories will approach straight lines, directed from the lens towards the focus of the wave. When the coincidence detections are considered, it will appear as if particle two departs from the place of detection of particle one.

This completes the Bohmian analysis of the phenomenon of optical imaging. Before the fragments reach the lens, they move along straight lines from the place of decay. Note that this place of decay is not fixed, in order to guarantee the momentum correlation ${\bf p}_1 + {\bf p}_2 = {\bf 0}$. When one of the particles reaches the lens, its direction of movement will change in accordance with the classical thin lense approximations. We assumed hereby the place of decay to be centered between the right detector and the lens. It can be expected, although it is not proven, that a random place of decay (for example near the lens) will lead to the same results in the Bohmian description of the experiment.  

\section{Conclusions}  
In conclusion we showed that Peres' analysis concerning the question to what extent opposite momenta lead to opposite directions, is only valid in the long distance regime. In the short time regime there is an additional source of angular deviation. On the other hand the statement ``opposite momenta lead to opposite directions'' is true in the Bohmian language. I.e. Bohmian particles travel in opposite directions when the wavefunction has eigenvalue zero for the total momentum operator (however from an unknown place of departure). We also showed that Bohmian trajectories can be used to gain insight into the phenomenon of ghost imaging as reported by Pittman {\em et al.} In fact Bohmian mechanics can be used to describe similar quantum optical experiments as well, such as Popper's experiment in the version of Kim and Shih \cite{kim99} and the experiment reported by Strekalov {\em et al.}\ \cite{Strekalov95}. In particular, Bohmian mechanics justifies the use of the conceptual pictures of photon trajectories present in these papers. Note that the experiment of Kim and Shih is very illustrative for the need for perfect momentum correlation of the photons, or equivalently that there must be very little restriction on the place of creation of the photons to create a perfect image. This is because Kim and Shih failed in their original intention to perform Popper's gedankenexperiment, due to the restricted diameter of the pump beam used in the experiment. The imperfect momentum correlation then led to an imperfect optical image \cite{short00}. This is immediately obvious when we consider our Bohmian picture of optical imaging, because if the momentum correlation is imperfect, the Bohmian particles will not move in opposite directions before the system reaches the lens. 
 
\appendix
\section*{Appendix}
Conservation of energy requires that the energy of the total system equals the energy of the decaying system $E_0$. If this decaying system is initially at rest, $E_0$ will be the rest energy of the system. Suppose now that we take a delta--distribution for this energy restriction i.e.\ 
\begin{equation}
F({\bf p}_1 , {\bf p}_2) =f({\bf p}_1,{\bf p}_2) \delta(E - E_0)
\label{50}
\end{equation}
where $f$ determines the momentum correlation (this is for example the distribution in Eq.~(\ref{10}) or Eq.~(\ref{24})). If we take a distribution $f$ which is real and symmetric, i.e.\ $f({\bf p}_1 , {\bf p}_2) = f(-{\bf p}_1 , -{\bf p}_2)$ then the probability currents of the two particles are both zero for all times, i.e.\
\begin{equation}
{\bf j}_i= \frac{\textrm{Im} (\psi^* {\bf \nabla}_{{\bf r}_i} \psi)}{m_i} = {\bf 0}, \qquad i=1,2.
\label{50.1}
\end{equation}
This implies that the particles show no evolution. In Bohmian mechanics this represents particles that stand still, because the Bohmian speeds are defined as $\ud {\bf R}_i / \ud t = {\bf j}_i/|\psi|^2 $.  As a result the considered distribution in Eq.~(\ref{50}) doesn't correspond to a decaying system. We can resolve this problem by allowing a finite energy width centered around $E_0$. However, it will follow that, if the wavefunction displays strong momentum correlation in the sense that ${\bf p}_1 + {\bf p}_2 = {\bf 0}$, the restriction to a small energy width only involves a minor broadening of the wavefunction, which implies that we can leave the restriction on the total energy aside for our qualitative analysis. 

For the momentum distribution $f$ we will take the distribution in Eq.~(\ref{24}), i.e.\ $f({\bf p}_1 , {\bf p}_2) \sim \de ({\bf p}_1 + {\bf p}_2)  e^{-\alpha p^2_1 / \hbar}$. The restriction on the total energy is accomplished by integrating over values for $({\bf p}_1 , {\bf p}_2)$ for wich $E_- \le E \le E_+$, for a certain minimum energy value $E_-$ and a certain maximimum energy value $E_+$. We therefore define the following function
\begin{displaymath}
 \textrm{disc}(E_{\pm})({\bf p}_1 , {\bf p}_2) = \left\{  \begin{array}{ll} 1 & \textrm{if} \quad  \frac{p^2_1}{2m_1} + \frac{p^2_2}{2m_2} \le E_{\pm} \\ 
0   & \textrm{otherwise} \end{array} \right. 
\label{51}
\end{displaymath}
The momentum distribution then becomes
\begin{equation}
F({\bf p}_1 , {\bf p}_2) =f({\bf p}_1,{\bf p}_2)\big[ \textrm{disc}(E_{+}) - \textrm{disc}(E_{-}) \big].
\label{52}
\end{equation}
The resulting wavefunction of the system is then
\begin{eqnarray}
\psi({\bf r}_1 , {\bf r}_2,t) &=& \int f({\bf p}_1,{\bf p}_2)\big[ \textrm{disc}(E_{+}) - \textrm{disc}(E_{-}) \big]  e^{i({\bf p}_1 \cdot {\bf r}_1 +  {\bf p}_2  \cdot {\bf r}_2 - Et)/\hbar}  \ud {\bf p}_1  \ud {\bf p}_2   \nonumber\\
   &\sim& \int e^{i{\bf p} \cdot ({\bf r}_1 - {\bf r}_2)/\hbar - (it/2\mu + \al)p^2/\hbar}  \big[ \textrm{disc'}(E_{+}) - \textrm{disc'}(E_{-}) \big] \ud {\bf p}
\label{53}
\end{eqnarray}
where
\begin{displaymath}
 \textrm{disc'}(E_{\pm})({\bf p}) = \left\{   \begin{array}{ll} 1 & \textrm{if} \quad p^2/2\mu \le E_{\pm} \\ 
0   & \textrm{otherwise} \end{array} \right. 
\label{54}
\end{displaymath}
If we write Eq.~(\ref{53}) as a Fourier transform then we can apply the convolution theorem
\begin{eqnarray}
\psi({\bf r}_1 , {\bf r}_2,t) &\sim& \mathcal{F}^+_{ \{( {\bf r}_1 -  {\bf r}_2)/\hbar \}} \big( e^{- (it/2\mu + \al)p^2/\hbar}\big) \otimes \mathcal{F}^+_{ (\{ {\bf r}_1 -  {\bf r}_2)/\hbar \}} \big(\textrm{disc'}(E_{+}) - \textrm{disc'}(E_{-})\big) \nonumber\\
&\sim& h(x,t)  \otimes g(x)
\label{55}
\end{eqnarray}
where we have defined
\begin{eqnarray}
h(x,t) &=& \mathcal{F}^+_{ \{( {\bf r}_1 -  {\bf r}_2)/\hbar \}} \big( e^{- (it/2\mu + \al)p^2/\hbar}\big), \nonumber\\
g(x) &=& \mathcal{F}^+_{ \{( {\bf r}_1 -  {\bf r}_2)/\hbar \}}\big(\textrm{disc'}(E_{+}) - \textrm{disc'}(E_{-})\big), \nonumber\\
x &=&  |{\bf r}_1 -  {\bf r}_2|.
\label{55.10}
\end{eqnarray}
We present now two methods to show that the restriction on the energy can be relinquished, without changing the qualitative analysis. The first method proceeds as follows. The first function in the convolution in Eq.~(\ref{55}) is just the wavefunction in Eq.~(\ref{25}),
\begin{equation}
h(x,t) \sim \bigg(\frac{\pi \hbar}{\al + it/2\mu} \bigg)^{3/2} e^{-({\bf r}_1 - {\bf r}_2)^2 / 4 \hbar (\al + \frac{it}{2\mu} )} .
\label{55.1}
\end{equation}
If we define $a_{\pm} =2 \pi \sqrt{E_{\pm} 2\mu}/ \hbar $ then the second function in the convolution in Eq.~(\ref{55}) in two--dimensional physical space becomes
\begin{equation}
g(x)  \sim \big[a_+ J_1 ( a_+ x  )- a_- J_1(a_- x )\big] /x  
\label{57}
\end{equation}
where $J_1$ is the first order spherical Bessel function. To evaluate $g(x)$ we will substitute some reasonable values for $E_+$ and $E_-$ in Eq.~(\ref{57}). In addition we will assume that the fragments have equal masses so that we can put $2 \mu = m$, with $m$ the mass of one fragment. For $E_+$ we will take one percent of the rest mass of the total system in order to avoid the relativistic regime, $E_+ = 0.02mc^2$. We will take an energy gap of $0.001E_+$, so that $E_- = 0.999E_+$. In this way $a_+ \approx 5.58309/\lambda_c$ and $a_- \approx 5.58023/\lambda_c$, where $\lambda_c$ is the Compton wavelength of the fragments. In Fig.\ 2 the function $g(x) = \big[a_+ J_1 ( a_+ x  )- a_- J_1(a_- x )\big] /x$ is plotted for $x$ in units of the Compton wavelength $\lambda_c$.
\begin{figure}[t]
\begin{center}
\epsfig{file=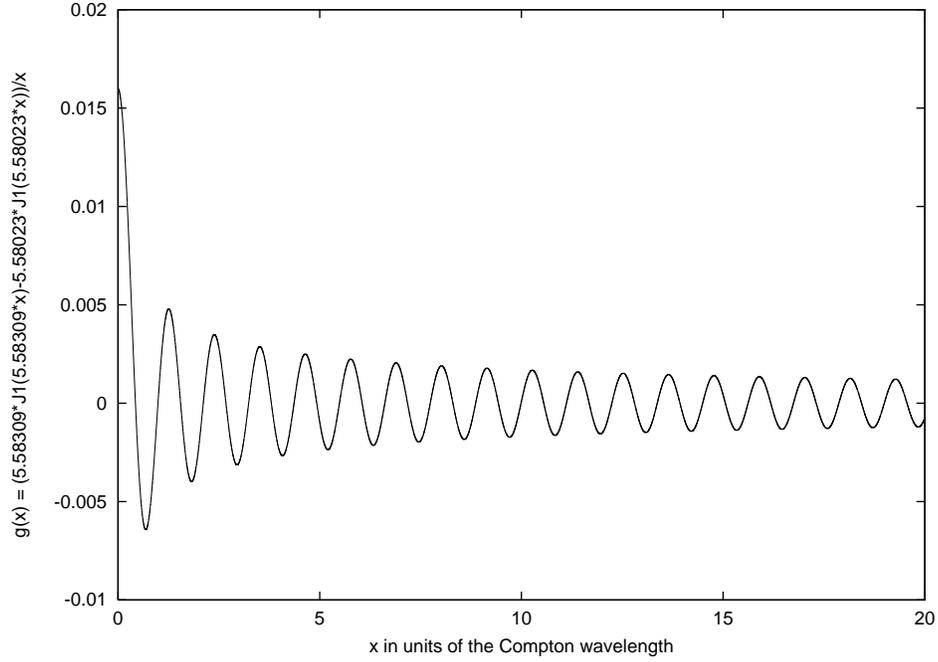}
\end{center}
\caption{The function $g(x) = \big[5.58309  J_1 ( 5.58309 x  )- 5.58023 J_1(5.58023  x )\big] /x$ is plotted for $x$ in units of the Compton wavelength $\lambda_c$.}
\end{figure}   

The figure shows that $g(x)$ is a rapidly oscillating function with a peak at $x = 0$. We now give the distribution of $h(x,t)$ at $t=0$ a width of the order of twenty times the Compton wavelength, which can be done by adjusting $\al$. Recall that the function $h$ was in fact the wavefunction of the system if we didn't restrict the energy. So the width of $h$ is in fact the measure of the initial nearness of the fragments, which is then of the order of twenty times the Compton wavelength. Then due to the rapid oscillation, the main contribution in the convolution will arise only from the peak in $g(x)$ at $x = 0$. This peak will result in only a small broadening of $h(x,0)$ so that $h(x,0) \otimes g(x)  \approx h(x,0)$. 

Because the width of $h(x,t)$ only increases with time, this approximation will become more precise with time. In conclusion we can put the unnormalised wavefunction equal to
\begin{equation}
\psi({\bf r}_1 , {\bf r}_2,t) \approx \bigg(\frac{\pi \hbar}{\al + it/2\mu} \bigg)^{3/2} e^{-({\bf r}_1 - {\bf r}_2)^2 / 4 \hbar (\al + \frac{it}{2\mu} )}.
\label{58}
\end{equation}
The larger the energy gap or the larger the rest energy, the more rapid the oscillation of $g(x)$ will be and the more peaked $g(x)$ will be at $x=0$, and as a result the more the approximation is valid.
 
A second way to achieve this result is to assume that $h(x,t)$ is very narrowly peaked at $x=0$ for $t=0$, so that $h(x,0) \approx \delta(x)$. This is the case if $\al$ approaches zero. As a result $h(x,0) \otimes g(x) \approx g(x)$. Thus in this case the unnormalised initial probability distribution is $g(x)^2$. This distribution is plotted in Fig.\ 3 with again $x$ in units of the Compton wavelength. This figure shows that most of the probability is concentrated within a few times the Compton wavelength.
\begin{figure}[t]
\begin{center}
\epsfig{file=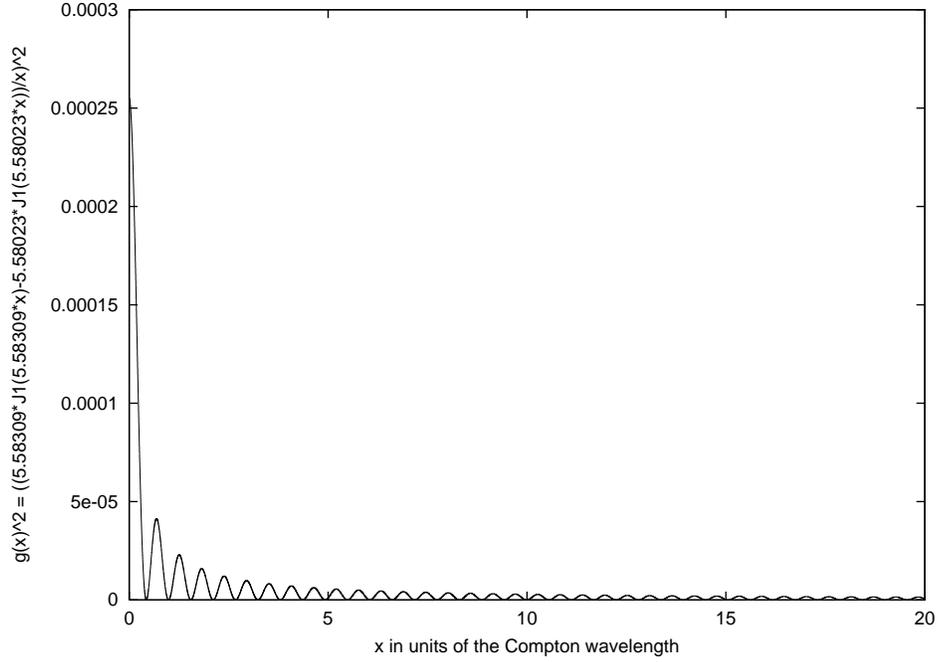}
\end{center}
\caption{The function $g(x)^2 = \frac{\big[5.58309 J_1 ( 5.58309 x)- 5.58023 J_1(5.58023 x )\big]^2}{x^2}$ is plotted for $x$ in units of the Compton wavelength $\lambda_c$.}
\end{figure}   

This implies that the particles depart from a very narrow region, only a few times the Compton wavelength in diameter, from each other. However, it is a priori unknown wether the Bohmian particles will travel along straight lines, in the direction of their connecting line, in this case . But this can be easily proven by invoking symmetry arguments. Because the wavefunction $\psi({\bf r}_1 , {\bf r}_2,t)$ in Eq.~(\ref{55}) is only dependent on the difference $|{\bf r}_1 - {\bf r}_2|$, the Bohmian velocities (as defined in Eq.~(\ref{27})) will be opposite and the Bohmian particles will travel along straight lines in opposite directions.

In conclusion, we have shown that the energy restriction does not put a restriction on the Bohmian picture of the system. The only effect of the energy restriction is a minor broadening of the probability density.

\section*{Acknowledgments}
WS acknowledges financial support from the F.W.O.\ Belgium. JDN acknow\-led\-ges financial support from IWT (``Flemish Institute for the scientific--tech\-no\-lo\-gi\-cal research in industry''). WS is especially thankful to A.\ Short for useful discussions and detailed explanation of his paper. The authors are also indebted to the anonymous referees who's remarks contributed to the development of the paper.

\end{document}